# Vibrational Properties of a Naturally Occurring Semiconducting van der Waals heterostructure

(Date: July 14th, 2021)

V.Z. Costa[1], Liangbo Liang[2], Sam Vaziri[3], Addison Miller[1], Eric Pop[3,4,5], and A. K. M. Newaz[1]

[1]Department of Physics and Astronomy, San Francisco State University, San Francisco, California 94132, USA

[2] Center for Nanophase Materials Sciences, Oak Ridge National Laboratory, Oak Ridge, Tennessee 37831, United States

[3]Department of Electrical Engineering, Stanford University, Stanford, California 94305, USA

[4]Department of Materials Science and Engineering, Stanford University, Stanford, California 94305, USA

[5]Precourt Institute for Energy, Stanford University, Stanford, California 94305, USA

**Abstract**

We present vibrational properties of Franckeite, which is a naturally occurring van der Waals heterostructure consisting of two different semiconducting layers. Franckeite is a complex layered crystal composed of alternating $SnS_2$-like pseudohexagonal and PbS-like pseudotetragonal layers stacked on top of each other, providing a unique platform to study vibrational properties and thermal transport across layers with mass density and phonon mismatches. By using micro-Raman spectroscopy and first-principles Raman simulations, we found that the PbS-like pseudotetragonal structure is mostly composed of $Pb_3SbS_4$. We also discovered several low-frequency Raman modes that originate from the intralayer vibrations of the pseudotetragonal layer. Using density functional theory, we determined all vibrational patterns of Franckeite, whose signatures are observed in the Raman spectrum. By studying temperature dependent Raman spectroscopy (300 K - 500 K), we have found different temperature coefficients for both pseudotetragonal and pseudohexagonal layers. We believe that our study will help understand the vibration modes of its complex heterostructure and the thermal properties at the nanoscale.

**INTRODUCTION**

Van der Waals (vdW) heterostructures obtained by stacking dissimilar two dimensional (2D) materials such as graphene (Gr), transition metal dichalcogenides (TMDs), or hexagonal boron nitride (*h*BN) has become a major research direction in condensed matter physics owing to novel electronic states and applications that may be realized in these vertical heterostructures.[1-6] These designer vdW heterostructures can demonstrate electronic, optical, and thermal properties that strongly differ from those of the constituent 2D materials, thus opening the door to obtain on-demand interesting physical, electrical, optical, optoelectrical behavior, and thermal properties, such as moiré exciton,[7, 8] and strongly-correlated quantum phenomena, including tunable Mott insulators at half-filling,[9, 10] unconventional superconductivity near integer-filling,[11-13] and ferromagnetism.[14] The standard fabrication method of all these heterostructures include manual or robotic vertical assembly of 2D stacks using deterministic placement methods.[5, 15, 16] Because of manual stacking, the neighboring layer interface may contain fabrication artifacts, such as foreign particles or bubbles between the interfaces, which may affect the measurement of intrinsic physical properties.[17-20] Hence, it is critical to measure intrinsic physical properties of interfaces with van der Waals heterostructures free of fabrication artifacts. We present here vibrational properties of a naturally grown van der Waals superlattices layered semiconductors, Franckeite, which is free from fabrication artifacts between semiconducting layers.[17-21]

Franckeite is a heterostructure composed of alternating sequences of weakly bound stacked PbS-like pseudotetragonal (Q) layers and $SnS_2$-like pseudohexagonal (H) layers attached by van der Waals interactions.[17-19] Thus, Franckeite (Fr) can be considered a naturally occurring vdW heterostructure analog of its lab-fabricated other 2D vdWHs. Recently, several groups have demonstrated exfoliation of Franckeite (mechanically and by liquid-phase exfoliation) down to the single unit cell and the exfoliated flakes have been assembled into electronic devices, energy conversion devices, and photodetectors operating in the near-infrared range.[17-19, 21]

Though Fr provides a fascinating 2D crystal heterostructure with possibilities of exploring new physics and developing many attractive applications, some of its fundamental properties remain unknown. First, though the crystal structure of H-layer is known ($SnS_2$), the composition crystal structure of Q-layer is highly debated. It is also unclear whether pseudotetragonal (*Q*) layer is predominantly made of $Pb_3SbS_4$ or it is Sn based $PbSnS_2$. Second, the Raman spectroscopy of Fr is not understood. Because Fr contains two different layered semiconducting materials, it may provide unique vibrational behavior because of mass density and phonon mismatches. Here, we answer these fundamental questions by performing micro-Raman spectroscopy, and first-principles density functional theory (DFT) studies. First, we determine the composition of the crystal structure of pseudotetragonal *Q*-layer. Second, we conduct extensive Raman spectroscopy of Fr and discover several low frequency Raman modes starting at 15 cm$^{-1}$. Third, we identify the vibrational modes responsible for the Raman-active behavior by using DFT calculations. Fourth, we perform the temperature dependent Raman spectroscopy of Fr vdW heterostructure for the first time to identify the thermal behavior of different semiconducting layers.

**RESULTS AND DISCUSSIONS**

We present our experimental and computational study of Raman spectroscopy of Franckeite immobilized flakes on $SiO_2$ (90 nm)/*p*$^+$Si substrate. The bulk Fr crystals were obtained from San Jose mine in Bolivia. We have prepared the sample by using micro-exfoliation from bulk samples. We point the reader to the work by Ray *et al.*, for a detailed characterization of the Franckeite by energy-dispersive X-ray (EDX) and scanning electron microscope.[19]

Fig.1 shows the optical and atomic force microscope (AFM) images of the micro-exfoliated sample. The flake $SiO_2$/ *p*$^+$Si substrate studied in this experiment is marked by an arrow as shown in Fig. 1(a). Fig.1 (b) shows the AFM image of the sample. The sample thickness is ~100 nm as seen in the height profile presented in the inset of Fig.1(b).

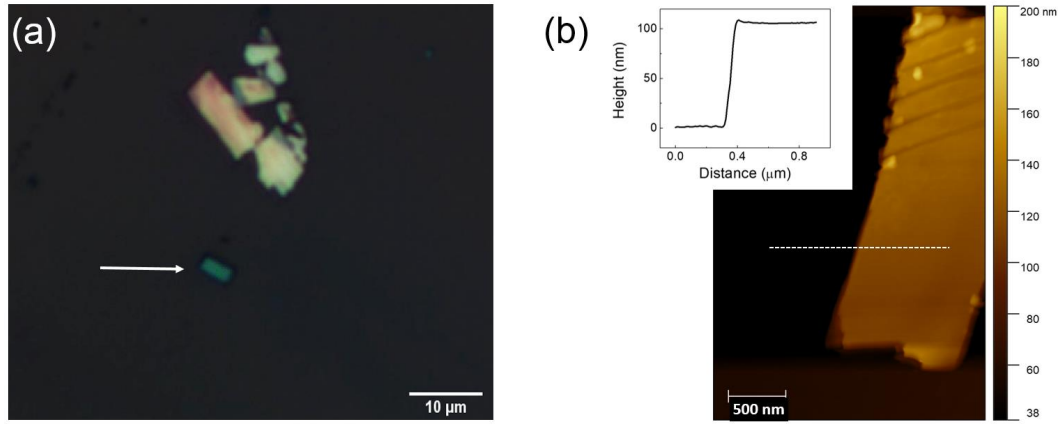

Figure 1: Optical and atomic force microscope image of a Franckeite sample. (a) The optical image of the sample used in the study (marked by an arrow). The sample is immobilized on a SiO$_2$ (90 nm)/$p^+$Si substrate (b) The AFM image of the same sample shown in (a). The sample thickness is ~100 nm. The inset is showing the height profile along the dashed line mark in the main panel.

Fig. 2(a) top panel presents the Raman spectrum of Fr measured at room temperature. We used a commercially available micro-Raman setup (Horiba LabRAM Evolution). We studied micro-Raman spectroscopy using both 532 nm and 633 nm excitation lasers and we observed that the Raman intensity is much larger with 633 nm laser than that by 532 nm laser. Differential reflectivity measurements confirmed higher absorption in the red range, suggesting resonance-enhanced Raman modes with 633 nm excitation source (see Supporting Information). In the rest of the study, we present our results with the 633 nm laser, which has a beam width ~ 1 μm. We have observed 10 clearly distinguishable peaks from the low frequency regime, as low as to 15 cm$^{-1}$, to the high frequency regime. The low frequency Raman modes are important to understand the crystal structure as well as the interlayer coupling.[22] To the best of our knowledge, we are reporting the low frequency Raman of Franckeite for the first time.

The Raman peaks near 140, 195, 255, 276, and 320 cm$^{-1}$ were reported by Velicky *et al.*, and us.[18, 19] It has been argued previously that the two peaks around 255 and 276 cm$^{-1}$ originate from the stibnite (Sb$_2$S$_3$) and the peak around 320 cm$^{-1}$ is from the berndtite (SnS$_2$) in the H-layer, respectively. The peaks at 43 and 199 cm$^{-1}$ may originate from the PbS lattices and SnS$_2$ lattices.[17-19] The origin of different Raman peaks, particularly the low frequency Raman modes we discovered, are still not completely understood, and their atomic vibrational patterns are not yet revealed. Hence a detailed study is necessary to elucidate the origin of different Raman peaks in Fr.

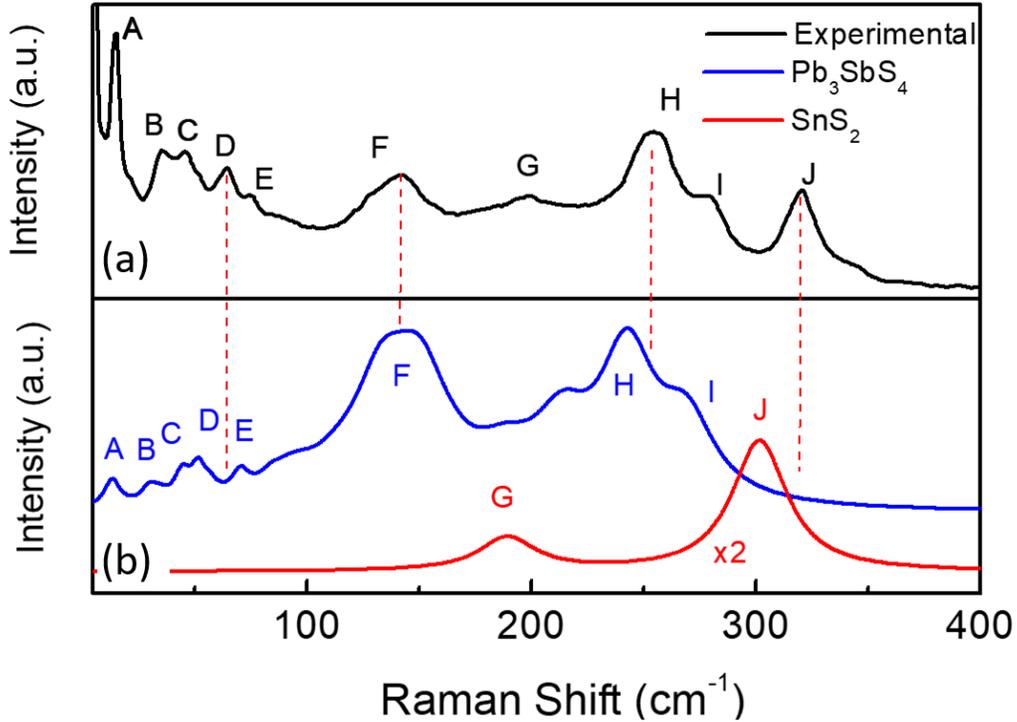

Figure 2: Raman spectroscopy of many-layer Franckeite. (a) Experimentally measured Raman spectroscopy at room temperature. The excitation laser wavelength is 633 nm. The Raman modes are marked alphabetically as A(15 cm$^{-1}$), B(35 cm$^{-1}$), C(45 cm$^{-1}$), D(64 cm$^{-1}$), E(75 cm$^{-1}$), F(140 cm$^{-1}$), G(195 cm$^{-1}$), H(255 cm$^{-1}$), I(276 cm$^{-1}$) and J(320 cm$^{-1}$). (b) Calculated Raman spectra of Franckeite are shown here. The blue line presents the Raman modes for $Pb_3SbS_4$ and the red line for $SnS_2$.

To determine the crystal structure of Franckeite by Raman spectroscopy, we performed first-principles DFT calculations using the plane-wave VASP package.[23-25] Individual $Pb_3SbS_4$, $PbSnS_2$, and $SnS_2$ layers were modeled for phonon and Raman scattering calculations, where $Pb_3SbS_4$ or $PbSnS_2$ is commonly considered as the structure of Q layer, while $SnS_2$ is considered as H layer in Franckeite (see Method Section for details).[17-19] We note that Q and H layers are incommensurate in Franckeite, and therefore it is computationally prohibitive for direct phonon and Raman calculations of the whole heterostructure. Since the interactions between the layers are weak, the Raman features of the heterostructure should be close to the composition of individual layers.

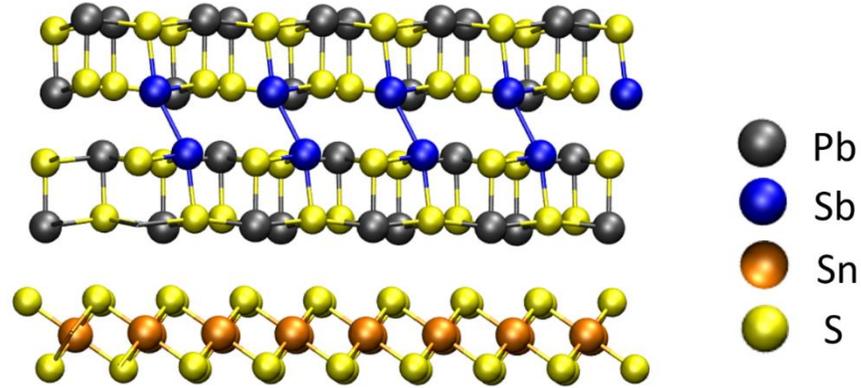

Figure 3: The crystal structure of Franckeite used for calculating the Raman spectra. The top is the Q-layer ($Pb_3SbS_4$) and the bottom is the H-layer ($SnS_2$). The colors assigned to different atoms are shown on the right.

According to our calculations, $Pb_3SbS_4$ layer has no symmetry (its crystal structure is shown in Figure 3 as part of Franckeite), so every phonon mode is Raman active in principle, which could explain why many Raman peaks are appearing in the experimental Raman spectra. This makes the phonon/Raman analysis challenging. By computing Raman intensities of $Pb_3SbS_4$, we can determine the phonon modes with strong Raman signals to compare with the experimentally measured Raman peaks. The $SnS_2$ layer, however, has high symmetry and it only has two distinctive Raman modes: one at 189 cm$^{-1}$ with $E_g$ symmetry, and the other one at 302 cm$^{-1}$ with $A_{1g}$ symmetry.

We have found that the experimentally measured Raman peaks are in good agreement with the calculated ones from $Pb_3SbS_4$ and $SnS_2$ layers (see Fig. 2 for comparison). The experimentally measured and first-principles calculated Raman peaks are also listed in Table 1. We also computed Raman peaks of $PbSnS_2$ and $SnS_2$ layers for comparison with the experimental data of Franckeite (Figure S1 in Supporting Information), where the mismatch can be clearly seen. For example, in the experimental Raman spectra, a peak I(279 cm$^{-1}$) appears as a right shoulder peak of peak H(255 cm$^{-1}$); however, $PbSnS_2$ layer does not exhibit such a distinctive feature and instead shows only one distinguishable peak around 271 cm$^{-1}$ (Figure S1), according to our calculations. In contrast, the computed Raman profile of $Pb_3SbS_4$ layer reproduces the experimentally observed shoulder peak feature (Fig. 2). These results suggest that Q layer is mostly like composed of $Pb_3SbS_4$ instead of $PbSnS_2$. Furthermore, in the low frequency region, our calculations also show that Raman peaks of $Pb_3SbS_4$ layer match better with experimental ones than those of $PbSnS_2$ layer. With the crystal structure of Q layer in Franckeite determined, we then computed the atomic vibrational patterns of all Raman peaks, including the low frequency and high frequency ones, and illustrate them in Fig.4.

| Low frequency modes Calculations (cm$^{-1}$) | Experiment (cm$^{-1}$) | High frequency modes Calculations (cm$^{-1}$) | Experiment (cm$^{-1}$) |
|---|---|---|---|
| 13 | 15 | 145 | 140 |
| 30 | 35 | 189 | 195 |
| 45 | 45 | 243 | 255 |
| 51 | 64 | 269 | 276 |
| 71 | 75 | 302 | 320 |

Table 1: Different Raman modes measured experimentally and determined by using first-principles calculations.

Finally, we acknowledge that there is still a discrepancy between the calculated and experimental Raman frequencies and intensities shown in Fig. 2 and Table 1. It is common that DFT calculations underestimate the phonon frequencies due to approximations adopted in the DFT methodology. The computed Raman intensities are not in exactly quantitative match with experimental ones owing to the adopted Placzek approximation and other factors missing in the calculations, such as the substrate effects, different laser polarization set-ups, resonant Raman effects, etc. In addition, individual Q and H monolayers are modeled in our DFT calculations instead of the realistic large vdW heterostructure with many layers due to the computational cost. Furthermore, the experimental structure contains substitutions of other elements such as Fe. Modeling of the exact experimental structure is very challenging since it requires a large supercell incorporating a small concentration of substitutional elements. Nevertheless, our calculated Raman spectra capture most of the experimental Raman features as seen in Fig. 2 and Table 1, such as the numerous low frequency Raman peaks, the distinctive shoulder peak feature formed between peak H and peak I, the broad peak F, the weak peak G, and the highest-frequency peak J from $SnS_2$ layer.

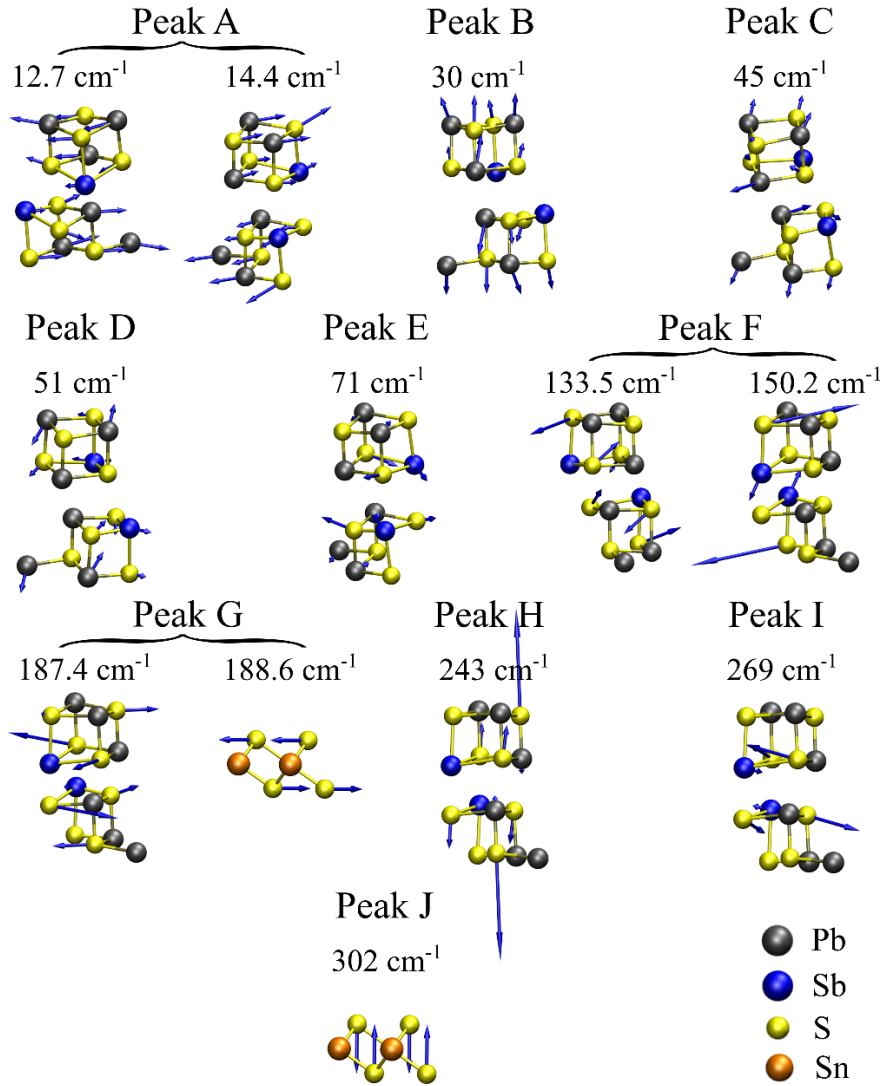

Figure 4: Calculated atomic vibrational patterns of experimental Raman peaks A-J in Franckeite. Raman modes from Q layer ($Pb_3SbS_4$) and H layer ($SnS_2$) are shown separately. The calculated phonon frequencies are shown for each mode. The symbols for different atoms are shown in the bottom right. Since $Pb_3SbS_4$ layer has no symmetry, every phonon mode is Raman active in principle, and thus there are many Raman peaks showing up in the experimental Raman spectra. Furthermore, some Raman peaks are contributed by multiple phonon modes close to each other (e.g., peak A, peak F, peak G, etc), which contributes to the large widths of Raman peaks in the experimental data. The $SnS_2$ layer has high symmetry and it only has two distinctive Raman modes: one around 188.6 $cm^{-1}$ (contributing to peak G), and the other one around 302 $cm^{-1}$ (corresponding to peak J). Other Raman peaks primarily originate from vibrations of $Pb_3SbS_4$ layer.

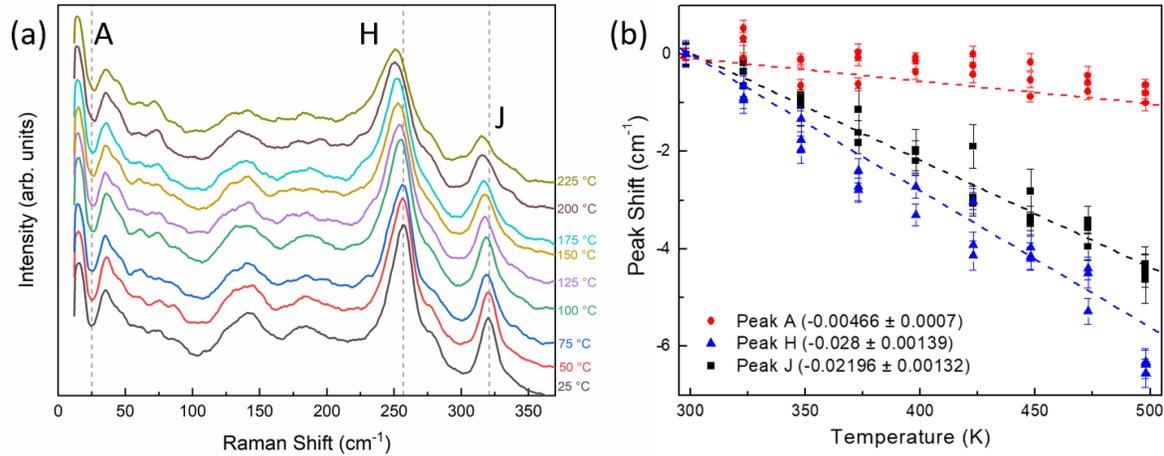

Figure 5: Temperature dependence of different vibrational modes. (a) The figures present the Raman spectra at different temperatures from 25 °C to 225 °C. All peaks are shifting to lower Raman shift values as we increase the temperature. (b) The peak shift positions (Δ= $\omega_{RT} - \omega_T$, where $\omega_{RT}$ and $\omega_T$ are the peak positions at room temperature and temperature $T$, respectively) of three representative Raman peaks; low frequency mode A, $Pb_3SbS_4$ peak H and $SnS_2$ peak J. We clearly see that all three peaks are evolving at different rates with respect to temperature.

To understand the effect of the Fr thickness on Raman modes, we performed room temperature Raman measurement of samples with a wide range of thickness from 6 nm to 100 nm. We have observed two important characteristics. First, we observed that the Raman peaks are independent of a wide range of Fr thickness from ~6 nm to ~100 nm. Second, we have not detected any Raman signal for samples below 6 nm.

Now we shift our focus to temperature dependent Raman spectroscopy. We have conducted Raman spectroscopy a wide range of temperatures ranging from 294 K (room temperature) to 500 K. The Raman spectra at different temperature is shown in Figure 5(a) revealing a redshift of the Raman peak as we increase the temperature. We observed that different Raman peaks shift at different rates.

Because the measured Raman peaks are broad, we determined the Raman peak position and errors using Gaussian fitting.[26] To enhance visualization of the temperature dependent Raman shift, we have calculated the peak shift, $\Delta = \omega_{RT} - \omega_T$, where $\omega_{RT}$ and $\omega_T$ are the peak positions at room temperature (~294 K) and temperature *T*, respectively. The peak shift, $\Delta$, of low frequency mode *A* peak, $Pb_3SbS_4$ peak H, and $SnS_2$ peak J are presented in Fig.5(b). We observed a higher rate of red-shift for H and J peaks compared to the low frequency *A* mode. To quantify the Raman shift as a function of temperature, we have used least-square fitting to determine the slope as $\omega = \omega_0 + \alpha T$. Here $\alpha$ is the first order temperature coefficient and $\omega_0$ is the Raman peak at the zero temperature.

We have measured the temperature coefficient $\alpha = 4.66 \times 10^{-3}, 28 \times 10^{-3}$ and $22 \times 10^{-3}$ cm$^{-1}$/K for A, H, and J peaks, respectively. Intriguingly, we note that the temperature coefficient for the low frequency peak is an order of magnitude lower than high frequency modes. Comparing to the reported values of other 2D layered materials, the temperature coefficient reported for the *G* peak in graphene, $A_{1g}$ mode in monolayer $WS_2$, and $A_{1g}$ in monolayer $MoS_2$ are, $16 \times 10^{-3}$, $12 \times 10^{-3}$ and $15 \times 10^{-3}$ cm$^{-1}$/K respectively.[27-29] The temperature coefficients of both *H* and *J*-peaks in Fr are larger than the temperature coefficients of the Raman peaks in graphene and $MoS_2$. The behavior of the rate of change in Raman peak position with temperature can vary for different phonon modes as well as for different materials.[30] The variation of the first order temperature coefficient of the Raman peak position of the normal modes is mainly due to the contribution from thermal expansion or volume contribution, and from the temperature contribution that results from anharmonicity.[30]

We have observed that the full width at half maximum (FWHM) increases significantly for both *F* and *G* peaks as shown in Fig.5(a), whereas there is a negligible FWHM increment for the low frequency mode as we increase the temperature. The peak height also decreases for all three peaks as the temperature increases (see Supplementary Information).

In conclusion, we have determined the composition of the crystal structure and vibrational properties of Fr by using Raman spectroscopy and first-principles Raman simulations. We have demonstrated the vibrational properties of Fr and how those phonon vibrations depend on temperature by measuring the temperature dependent Raman modes. Our study will help understand the atomic structure and vibrational modes of a natural heterostructure, and its heat transport management at the nanoscale.

**Methods:**

**Sample Fabrication:** Fr samples immobilized on different substrates were prepared by micro-exfoliation using polyimide tape of 0.015 mm of thickness and an adhesive layer of 0.06 mm of thickness, followed by characterization using optical microscopy. Raman spectroscopy and atomic force microscopy (AFM) were performed to verify the franckeite deposition and determine the sample thickness. The suspended Fr samples were picked-up and dropped off via dry-transfer technique with PET stamp on TEM 2000 (aperture size ~7.5 μm) mesh grids attached to a 90 nm SiO2 substrate with double-sided Kapton tape.

**Raman Characterization:** Confocal micro-Raman measurements were performed using commercial equipment (Horiba LabRAM Evolution). A long working-distance 100× objective lens with a numerical aperture of 0.6 was used. The excitation source was a 633 nm laser of power

~200 μW. The Raman spectra are measured using a grating with 1200 g/mm blazed at 500 nm and a solid-state-cooled CCD detector. To measure the temperature dependent Raman, we calibrated the temperature of the sample with a Linkam THMS600 stage in ambient air using the Si Raman peak.

**DFT Calculations:** We performed first-principles density functional theory (DFT) calculations using the plane-wave VASP package, where projector augmented wave (PAW) pseudopotentials were used for electron-ion interactions[23] and the Perdew-Burke-Ernzerhof (PBE) functional was for exchange-correlation interactions.[24, 25] Van der Waals (vdW) interactions were included using the DFT-D3 method. Individual $Pb_3SbS_4$, $PbSnS_2$, and $SnS_2$ layers were modeled for phonon and Raman scattering calculations, where $Pb_3SbS_4$ or $PbSnS_2$ is commonly considered as the structure of the *Q* layer in Franckeite, while $SnS_2$ is considered as the *H* layer in Franckeite.[17-19] Note that Franckeite, a naturally occurring vdW heterostructure, exhibits a non-commensurate layer match between the *Q* and *H* layers, and thus it would be computationally too costly for direct phonon and Raman calculations of the whole Franckeite heterostructure. Because the interactions between the layers are weak, the phonon and Raman features of the Franckeite heterostructure can be described by its individual component layers, i.e., $Pb_3SbS_4$ (or $PbSnS_2$) and $SnS_2$. Single-layer $Pb_3SbS_4$, $PbSnS_2$ and $SnS_2$ were modeled by a periodic slab geometry, where a vacuum separation of at least 21 Å in the out-of-plane direction (i.e., *z* direction) was set to avoid spurious interactions between periodic images. For the 2D slab calculations, a 12×12×1 k-point sampling was used for $Pb_3SbS_4$ and $PbSnS_2$, while a 24×24×1 k-point sampling was for $SnS_2$. For all systems, the cutoff energy was chosen as 350 eV, and all atoms were relaxed until the residual forces were below 0.001 eV/Å. Note that the in-plane lattice constants were optimized using the method of fixing the total volume (ISIF = 4 in VASP) to avoid the collapse of the vacuum separation in the *z* direction.

Based on the fully relaxed structures, phonon calculations were carried out using the finite difference scheme implemented in the Phonopy software to obtain phonon frequencies and eigenvectors.[31] Hellmann-Feynman forces in the supercell (2×2×1 for $Pb_3SbS_4$ and $PbSnS_2$, while 3×3×1 for $SnS_2$) were computed by VASP for both positive and negative atomic displacements (δ = 0.03 Å) and then used in Phonopy to construct the dynamic matrix, whose diagonalization provides phonon frequencies and phonon eigenvectors (i.e., vibrations). Raman scattering calculations were then performed within the Placzek approximation using the in-house developed Raman modeling package.[32-34] For the $j$-th phonon mode, Raman intensity is $I \propto \frac{(n_j+1)}{\omega_j}|e_i \cdot \tilde{R} \cdot e_s^T|^2$, where $e_i$ and $e_s$ are the electric polarization vectors of the incident and scattered lights respectively, and $\tilde{R}$ is the Raman tensor of the phonon mode.[35] $\omega_j$ is the frequency of the $j$-th phonon mode, and $n_j = (e^{\hbar\omega_j/k_BT} - 1)^{-1}$ is its Boltzmann distribution function at the given temperature $T = 300$ K. The matrix element of the (3×3) Raman tensor $\tilde{R}$ of the $j$-th phonon mode is[32, 33, 35]

$$\tilde{R}_{\alpha\beta}(j) = V_0 \sum_{\mu=1}^{N} \sum_{l=1}^{3} \frac{\partial \chi_{\alpha\beta}}{\partial r_l(\mu)} \frac{e_l^j(\mu)}{\sqrt{M_\mu}},$$

where $\chi_{\alpha\beta} = (\varepsilon_{\alpha\beta} - \delta_{\alpha\beta})/4\pi$ is the electric polarizability tensor related to the dielectric tensor $\varepsilon_{\alpha\beta}$, $r_l(\mu)$ is the position of the μ-th atom along the direction $l$, $\frac{\partial \chi_{\alpha\beta}}{\partial r_l(\mu)}$ is the derivative of the polarizability tensor (essentially the dielectric tensor) over the atomic displacement, $e_l^j(\mu)$ corresponds to the displacement of the μ-th atom along the direction $l$ in the $j$-th phonon mode (i.e., the eigenvector of the dynamic matrix), $M_\mu$ is the mass of the μ-th atom, and $V_0$ is the unit cell volume. For both positive and negative atomic displacements (δ = 0.03 Å) in the unit cell, the dielectric tensors $\varepsilon_{\alpha\beta}$ were computed by VASP[36] and then their derivatives were obtained via the finite difference scheme. Based on the phonon frequencies, phonon eigenvectors and the derivatives of dielectric tensors, Raman tensor $\tilde{R}$ of any phonon mode can be obtained, which subsequently yields the Raman intensity $I(j)$ under a given laser polarization set-up. Finally, based on the calculated Raman intensities $I(j)$ and phonon frequencies $\omega_j$, the Raman spectrum can be obtained after Lorentzian broadening.

**Acknowledgement**

V.Z.C., and A.K.M.N. acknowledge the support from the Department of Defense Award (ID: 72495RTREP). A.K.M.N. also acknowledge the support from the National Science Foundation Grant ECCS-1708907. All AFM measurements were supported by NSF for instrumentation facilities (NSF MRI-CMMI 1626611). All Raman spectroscopy data were acquired at the Stanford Nano Shared Facilities (SNSF), supported by the National Science Foundation under award ECCS-2026822. A portion of this research (Raman scattering modeling) used resources at the Center for Nanophase Materials Sciences, which is a U.S. Department of Energy Office of Science User Facility. L.L. acknowledges computational resources of the Compute and Data Environment for Science (CADES) at the Oak Ridge National Laboratory, which is supported by the Office of Science of the U.S. Department of Energy under Contract No. DE-AC05-00OR22725.

*Supporting Information:*

# Vibrational Properties of a Naturally Occurring Semiconducting van der Waals heterostructure

V.Z. Costa[1], Liangbo Liang[2], Sam Vaziri[3], Addison Miller[1], Eric Pop[3,4,5], and   A. K. M. Newaz[1]

[1]Department of Physics and Astronomy, San Francisco State University, San Francisco, California 94132, USA

[2] Center for Nanophase Materials Sciences, Oak Ridge National Laboratory, Oak Ridge, Tennessee 37831, United States

[3]Department of Electrical Engineering, Stanford University, Stanford, California 94305, USA

[4]Department of Materials Science and Engineering, Stanford University, Stanford, California 94305, USA

[5]Precourt Institute for Energy, Stanford University, Stanford, California 94305, USA

S1: Raman spectroscopy for $Pb_3SbS_4$

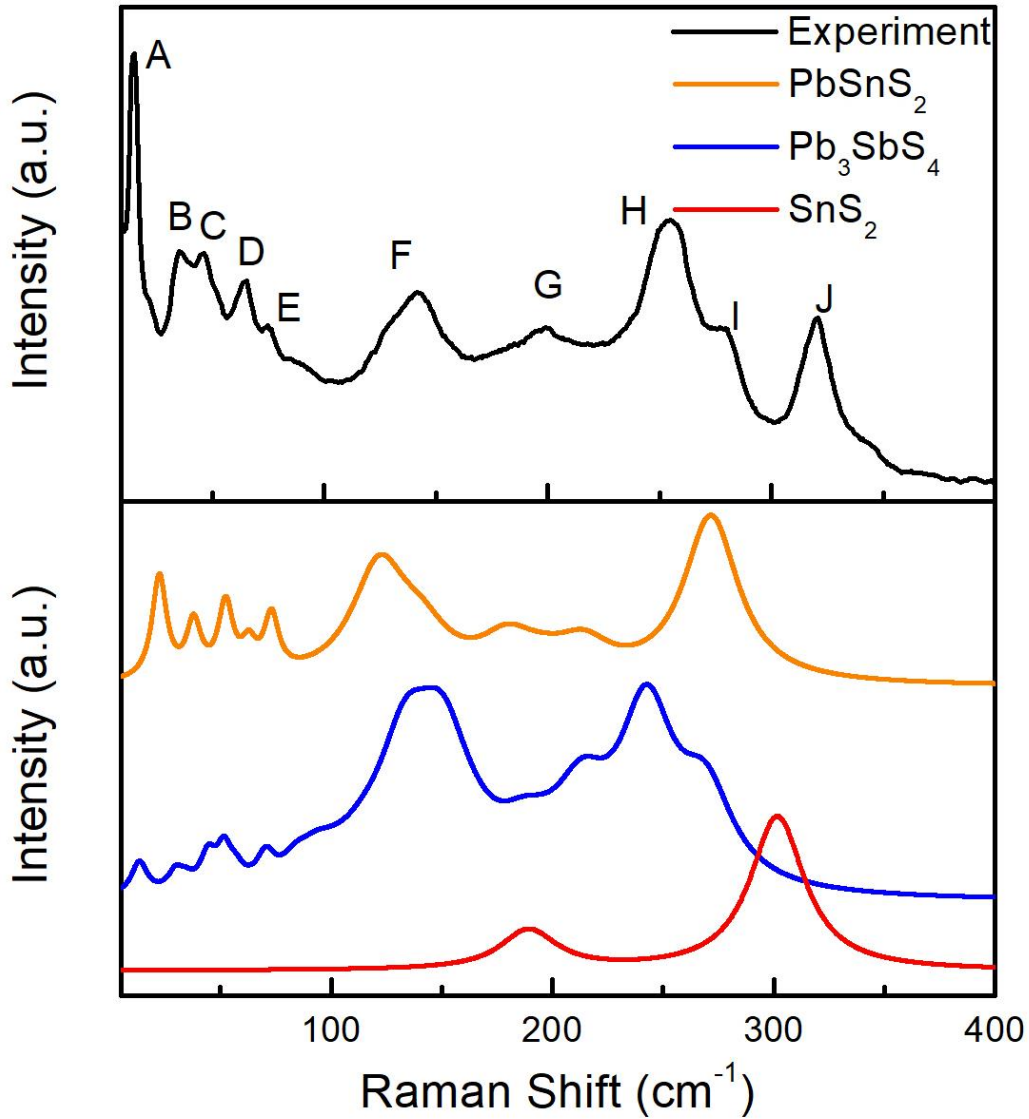

Figure S1: Raman spectroscopy of many layer Franckeite. (a) The plot presents the experimentally measured Raman spectroscopy at room temperature. The excitation laser wavelength is 633 nm. We have observed high frequency Raman modes as well as low frequency Raman modes. The Raman modes are marked alphabetically as A(14 cm$^{-1}$), B(35 cm$^{-1}$), C (45 cm$^{-1}$), D(64 cm$^{-1}$), E(75 cm$^{-1}$), F(140 cm$^{-1}$), G(195 cm$^{-1}$), H(255 cm$^{-1}$), I(276 cm$^{-1}$) and J(320 cm$^{-1}$). (b) Calculated Raman spectra of Franckeite are shown here. The upper (orange), middle (blue) and lower (red) plots present the Raman modes for $PbSnS_2$, $Pb_3SbS_4$ and $SnS_2$, respectively.

## S2: Differential reflectivity measurements of Franckeite

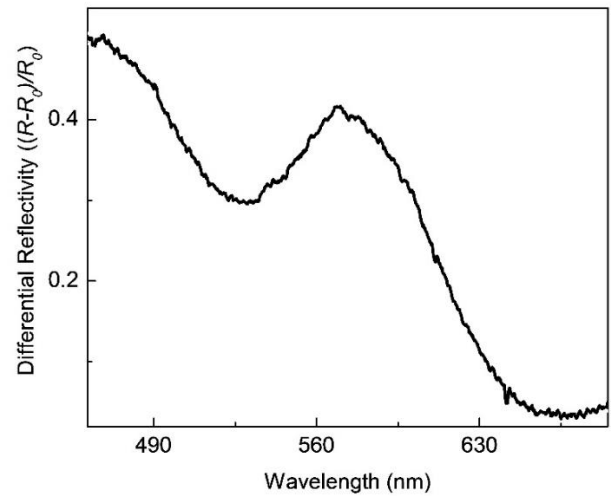

Figure S2: Differential reflectivity measurements of Franckeite. The differential reflectivity ($\frac{R-R_0}{R_0}$, where $R$ is the change in the reflectance signal from the sample and $R_0$ is from off the sample location. We have used a broad band thermal light source to illuminate the sample. The reflectance signal was measured by a spectrometer with 300 lines/mm grating.